\documentclass[prb,aps,twocolumn,showpacs]{revtex4}

\usepackage{epsfig}

\begin{document}

\title{$^{75}$As NMR study of single crystals of the heavily overdoped pnictide superconductors Ba$_{1-x}$K$_x$Fe$_2$As$_2$ ($x$ = 0.7 and 1)}

\author{S. W. Zhang $^{1}$}
\author{L. Ma$^{1} $}
\author{Y. D. Hou ${^1}$}
\author{J. S. Zhang $^{2}$}
\author{T. L. Xia ${^1}$}
\author{G. F. Chen $^{1}$}
\author{J. P. Hu$^{3}$}
\author{G. M. Luke$^{4,5}$}
\author{W. Yu$^{1}$}
\email{wqyu_phy@ruc.edu.cn}
\affiliation{
$^1$Department of Physics, Renmin University of China, Beijing 100872, China\\
$^2$School of Energy and Power Engineering, North China Electric Power University, Beijing 102206, China\\
$^3$Department of Physics, Purdue University, West Lafayette, IN 47907, USA\\
$^4$Department of Physics and Astronomy, McMaster University, Hamilton, Ontario L8S 4M1, Canada\\
$^5$Canadian Institute for Advanced Research, Toronto, Canada\\
}
\date{\today}
\pacs{74.70.-b, 76.60.-k}

\begin{abstract}

We performed $^{75}$As NMR studies on two overdoped high-quality Ba$_{1-x}$K$_{x}$Fe$_2$As$_2$ (x=0.7 and 1.0) single crystals.
In the normal states, we found a dramatic increase of the spin-lattice relaxation ($1/^{75}T_1$) from the x=0.7 to the x=1.0 samples. In KFe$_2$As$_2$, the ratio of 
$1/^{75}T_1TK_n^2$, where $^{75}K_n$ is the Knight shift, increases as temperature drops. These results indicate the existence of a new type of spin fluctuations in 
KFe$_2$As$_2$ which is accustomed to being treated as a simple Fermi liquid. In the superconducting state, we observe a step-like feature in the temperature dependence of 
the spin-lattice relaxation of the x=0.7 sample, which supports a two-gap superconductivity as the underdoped materials. However, the temperature scalings of $1/^{75}T_1$ 
below Tc in the overdoped samples are significantly different from those in the under or optimal doped ones. A power-law scaling behavior $1/^{75}T_1T\sim T^{0.5}$ is 
observed, which indicates universal strong low energy excitations in the overdoped hole-type superconductors.

\end{abstract}

\maketitle

The discovery of superconductivity at 26K in LaFeAsO$_{1-x}$F$_x$  \cite{Hosono_Jacs1} and the improvement of superconducting transition temperature above 50K in other 
iron pnictides \cite{Chen_XH, Chen_GF, Ren_ZA} have caused renewed interests in high-temperature superconductivity. All iron arsenides have a layered structure, where the 
FeAs plane is believed to be essential to the electronic properties. The Fermi surface of the parent compounds consists of two hole-pockets around the $\Gamma$ point, and 
one-hole pocket and two electron-pockets around the $M$ point \cite{Lebegue, Singh,Xu}. Upon either electron or hole doping, most compounds evolve from an 
antiferromagnetically ordered state  to a superconducting state \cite{Chen_GF}. The study of the interplay and the doping dependence of the band structure, the magnetism, 
and the pairing symmetry are crucial to understanding the mechanism of
superconductivity.

The pairing symmetry of Fe-based superconductors has not been fully established. A promising candidate is  the so called $S_\pm$ pairing symmetry which has opposite sign 
on the electron and hole pockets. This pairing symmetry has been argued in both weak coupling and strong coupling theoretical models\cite{Singh1,seo2008, Kuroki1, 
LeeDH1}.  The existence of both electron and hole pockets are critical in producing the $S_\pm$ pairing symmetry. In particular, the weak coupling model  is entirely 
based on the interband interactions between the electron and hole pockets. Experimentally, one heavily studied series of materials is the hole-doped 
Ba$_{1-x}$K$_x$Fe$_2$As$_2$ \cite{Rotter, Chen_XH2}. The ARPES study showed two isotropic s-wave superconducting gaps in the optimal-doped Ba$_{0.6}$K$_{0.4}$Fe$_2$As$_2$ 
\cite{Ding2}. However, the relative sign between two pockets has not been measured. From NMR, the absence of a coherence peak and the power-law-like behavior of the 
spin-lattice relaxation rate (SLRR) below Tc of the underdoped and the optimal-doped Ba$_{1-x}$K$_x$Fe$_2$As$_2$ materials  \cite{Zheng2, Fukazawa2, Yashima1,Julien} 
rules out a conventional single band s-wave, although it is still unable to differentiate the S$_\pm$, the d-wave or other nodal pairing symmetries \cite{Zheng2, 
Fukazawa2, Yashima1, Nagai1, Parish2008}.

Upon hole doping,  a shrinkage of the electon-pockets is seen in Ba$_{1-x}$K$_x$Fe$_2$As$_2$ by ARPES\cite{Ding2, Ding3}. In particular in KFe$_2$As$_2$ (x=1.0),  the 
electron pockets disappear completely \cite{Ding3}. Since the weakening of the interband interaction is expected upon heavy doping, it raises the question if the 
magnetism and the pairing mechanism change. Distinctive properties have also been reported in KFe$_2$As$_2$ comparing to low dopings. For example, T$_C$ remains finite at 
about 3K \cite{Chu_PRL, Chen_XH2, Terashima1}, and the H$_{C2}$ is anisotropic \cite{Terashima1} rather than isotropic \cite{Yuan_nature}. Strong low energy excitations 
below T$_C$ is observed by NQR which supports a multiple gap superconductor \cite{Fukazawa1}. In order to understand the evolution of the normal state properties and the 
pairing symmetry in the overdoped side, it is essential to study the properties on high quality crystals and also with more intermediate dopings.

In this work, we performed NMR studies on high-quality overdoped Ba$_{0.3}$K$_{0.7}$Fe$_2$As$_2$ and KFe$_2$As$_2$ single crystals. Surprisingly,  we found spin 
fluctuations are strongly enhanced in the normal state SLRR of the heavily overdoped sample KFe$_2$As$_2$.  A deviation from a simple Fermi liquid is indicated by the 
decrease of $1/T_1TK_n^2$ with temperature in the normal state of KFe$_2$As$_2$, where $K_n(T)$ is the Knight shift. Below T$_C$, we found a clear step-like feature in 
the SLRR in Ba$_{0.3}$K$_{0.7}$Fe$_2$As$_2$, which supports a two-gap superconductivity. More importantly, the power-law behavior of the SLRR as a function of 
temperature, $1/T_1\sim T^{\alpha _s}$, has a universal power law exponent $\alpha_s\sim 1.5$ below T$_C$ in the overdoped samples, in contrast to the large and 
non-universal power-law exponent $\alpha _s$ observed in the under- or optimally doped samples\cite{Zheng2, Fukazawa2, Yashima1}.

Our Ba$_{0.3}$K$_{0.7}$Fe$_2$As$_2$ and KFe$_2$As$_2$ single crystals were grown by the FeAs-flux method \cite{GFC2}. The NMR crystals were plate-like with a surface area 
of 2.5mm*1.2mm. We mainly perform the $^{75}$As measurements with different field strength and orientations. The SLRR is measured by the inversion-recovery method on the 
central transition, and the recovery curve of the SLRR is fitted with a standard double-exponential form of an S=3/2 spin, $1-\frac {m(t)}{m(0)}=0.9exp(\frac 
{-6t}{T_1})+0.1exp(\frac{-t}{T_1})$. The usage of single crystals also enables us to measure the Kinght shift and the anisotropy of $1/T_1$ accurately.

\begin{figure}
\includegraphics[width=6cm, height=5cm]{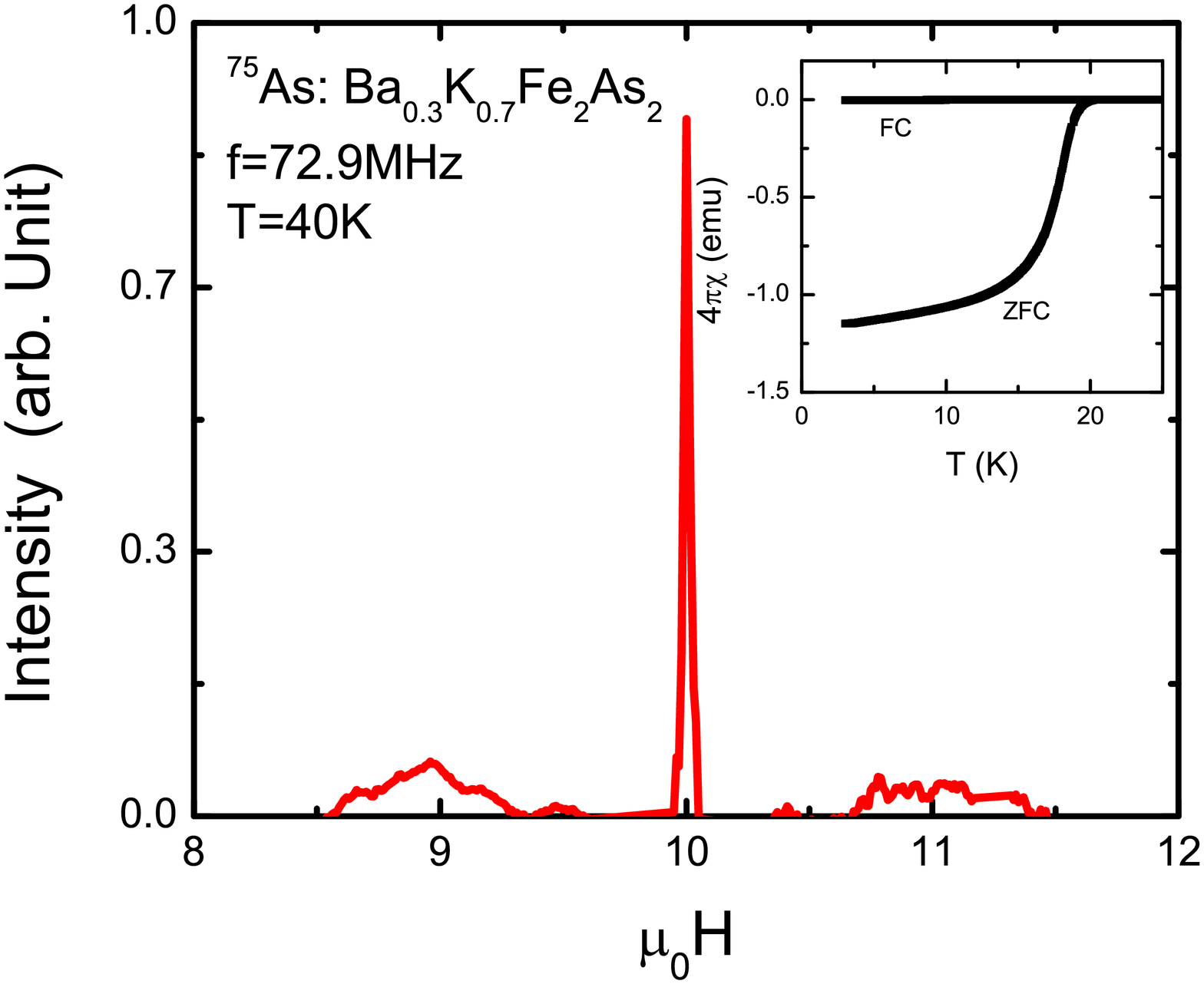}
\caption{\label{IHx70}(Color Online) The $^{75}$As NMR spectrum of the Ba$_{0.3}$K$_{0.7}$Fe$_2$As$_2$ single crystal at a fixed frequency 72.9MHz. Inset: The dc 
susceptibility of the crystal under field cooling (FC) and zero field cooling (ZFC) with a 100Oe field.}
\end{figure}

In Fig.~\ref{IHx70}(inset), the magnetic susceptibility of the Ba$_{0.3}$K$_{0.7}$Fe$_2$As$_2$ single crystal is shown with a 100Oe field. From the ZFC data, the sample 
is 100$\%$ superconducting in volume, and bulk superconductivity starts around 19K, and the major transition completes around 15K. The $^{75}$As (S=3/2) NMR spectra of 
the crystal at a fixed frequency 72.9MHz with field along the crystal c-axis is shown in Fig.~\ref{IHx70}, with a center transition at  $\mu _0H\sim 10T$ and two 
satellites at 8.8T and 11.2T respectively. In the following, we focus on the SLRR of the central transition (spin 1/2$\rightarrow$-1/2 transition).

\begin{figure}
\includegraphics[width=7cm, height=6cm]{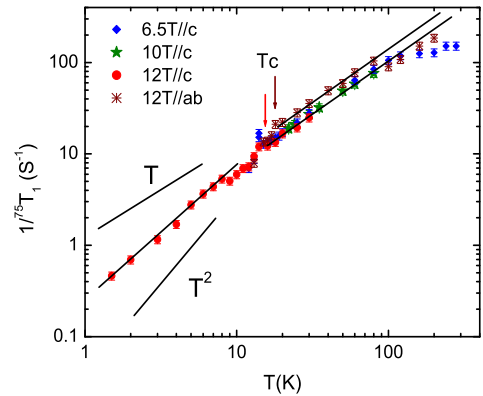}
\caption{\label{T1Asx70}(Color Online) The temperature dependence of $^{75}$As spin lattice relaxation rate of the Ba$_{0.3}$K$_{0.7}$Fe$_2$As$_2$ single crystal measured 
at four different fields, 12T//c, 10T//c, 6.5T//c and 12T//ab.}
\end{figure}

In Fig.~\ref{T1Asx70}, we show the $1/^{75}T_1$ of the Ba$_{0.3}$K$_{0.7}$Fe$_2$As$_2$ crystal at various field with two orientations (H//c and H//ab). In the normal 
state, $1/^{75}T_1$ can be fit by a power-law $1/^{75}T_1\sim T^{\alpha _n}$, with $\alpha _n\approx 1.1$ from T$_C$ up to about 100K, which is close to the Fermi liquid 
Korringa relation.  The normal-state SLRR is anisotropic with $1/T_1^{ab}$ larger than $1/T_1^c$ for two different field orientations, and the anisotropic factor, 
estimated to be $T_1^c/T_1^{ab} \approx 1.3$, holds for all temperatures above T$_C$. From 100K and above, both the transport \cite{Chen_XH2} and our $1/^{75}T_1$ deviate 
from a simple Fermi liquid behavior. This feature has also been observed in previous NMR measurements of Ba$_{1-x}$K$_{x}$Fe$_2$As$_2$ with low different doping
concentrations\cite{Zheng2}.  The origin of such a deviation is not completely understood.

\begin{figure}
\includegraphics[width=7cm, height=6cm]{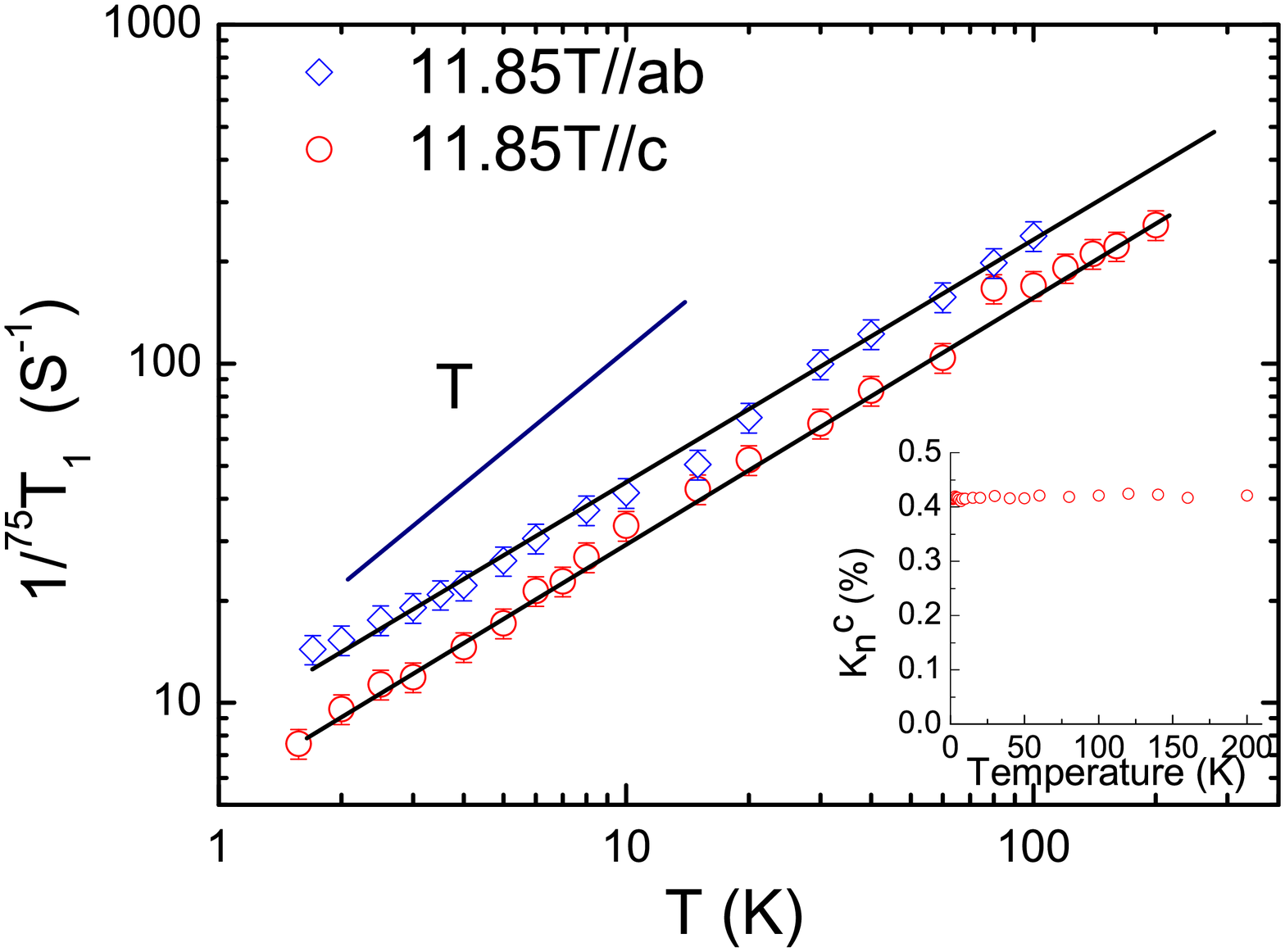}
\caption{\label{T1Asx100}(Color Online) The spin-lattice relaxation rate $1/^{75}T_1$ of the KFe$_2$As$_2$ under 12T field with two orientation H//c and H//ab. Inset: The 
Knight shift $K_n^s$ of the sample with H=11.85T//C.}
\end{figure}

We took similar measurements on KFe$_2$As$_2$. The superconducting transition temperature of our sample is   about 3K from resistivity, close to earlier reports 
\cite{Chu_PRL, Chen_XH2, Terashima1}.  The $^{75}$As spectrum is very narrow, with a FWHM only 16KHz under 12T//c field at 200K, which indicates high quality of the 
sample. We studied the SLRR of $^{75}As$ by NMR under 12T field with H//c and H//ab, as shown in Fig.~\ref{T1Asx100}. As both fields are much higher than the H$_{C2}$ 
\cite{Terashima1}, we were able to measure the normal state SLRR down to 1.5K. The comparison between two different field orientations gives the
anisotroic factor of the SLRR $T_1^{c}/T_1^{ab}\approx 1.4$.

We first discuss the normal state properties of both dopings. Our low-temperature resistivity shows a T$^2$ dependence, or a Fermi-liquid-like behavior, for both samples, 
which is consistent with other reports \cite{Chen_XH2}. However, our NMR data also suggest the existence of strong spin fluctuations and indicates a  deviation from a 
simple Fermi liquid behavior in x=1.0 within the following several aspects.

First, the temperature dependence of the SLRR in the x=1.0 sample deviates from the Fermi liquid behavior. The $1/T_1$ is fit by a power law $1/^{75}T_1\sim T^{\alpha 
_n}$ with the exponent $\alpha _n\approx 0.75$, as shown in Fig.~\ref{T1Asx100}, over two decades of temperature above 1.5K for both field orientations. Similar power-law 
behavior and power-law exponent $\alpha _n\approx 0.8$ were reported by a NQR measurement on KFe$_2$As$_2$ powders\cite{Fukazawa1}. The change of the Knight shift, as 
shown in Fig.~\ref{T1Asx100} (inset), is negligibly small up to 200K.  Taking account the power-law behavior of the SLRR, the ratio $1/TT_1K_n^2(T)$ does not follow the 
modified Korringa relation ($1/TT_1K_n^2(T)\sim const$). Instead, $1/T_1TK_n^2(T)\propto T^{-0.25}$ increases as temperature drops, which is a signature of spin 
fluctuations.

\begin{figure}
\includegraphics[width=6cm, height=5cm]{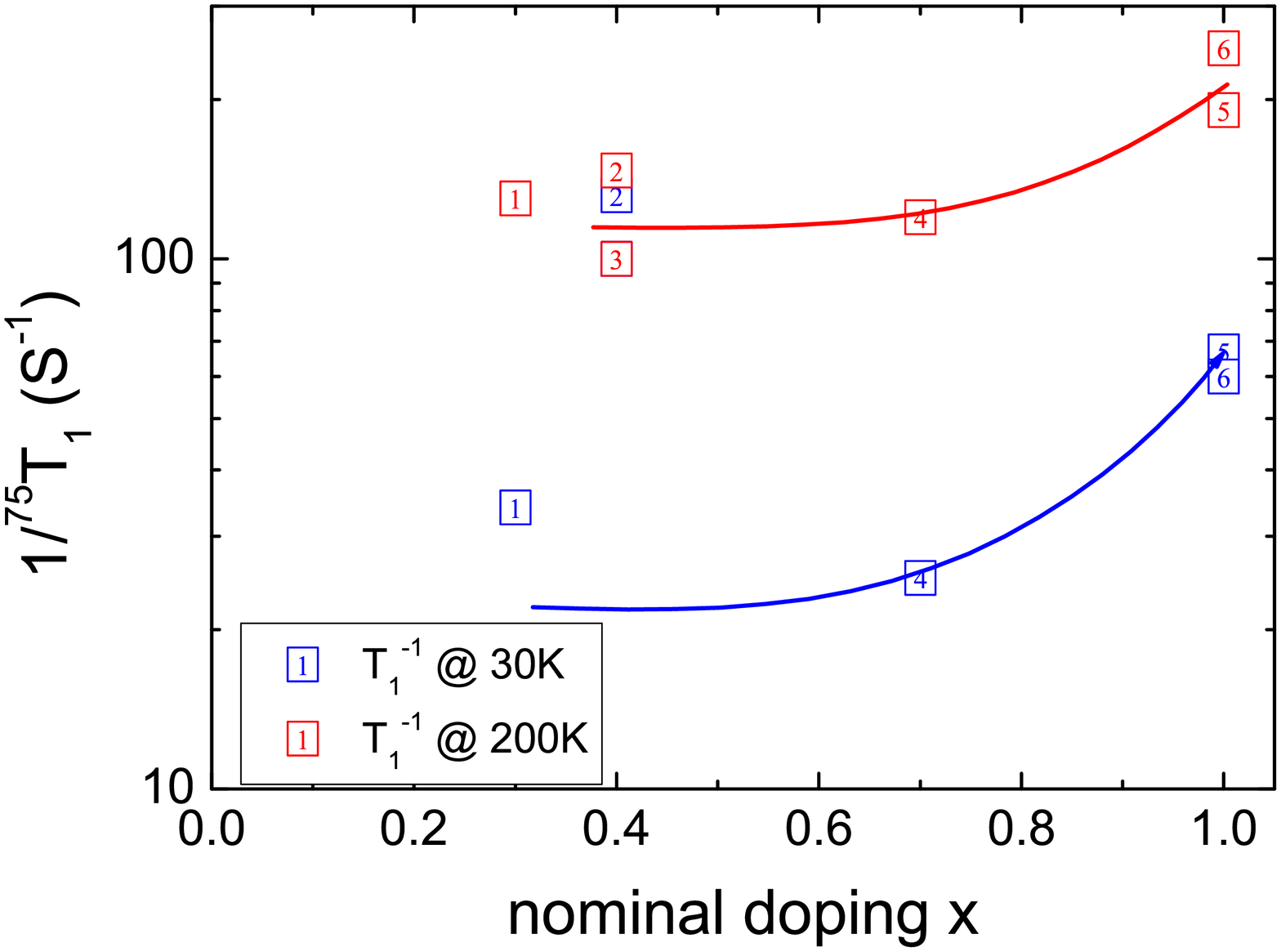}
\caption{\label{invT1x}(Color Online) The normal state spin-lattice relaxation rate $1/^{75}T_1^c$ at 30K and 200K in 1: x=0.3 (Ref. \cite{Zheng2}), 2: x=0.4 (Ref. 
\cite{Yashima1}), 3: x=0.4 (Ref. \cite{Fukazawa2}),  4: x=0.7 (current work), 5: x=1.0 (current work) and 6: x=1.0 (NQR, Ref. \cite{Fukazawa1}) 
Ba$_{1-x}$K$_{x}$Fe$_2$As$_2$ samples.}
\end{figure}

Second, our NMR data show a prominent change of the spin dynamics with doping close to x=1.0. In Fig.~\ref{invT1x}, the normal-state SLRR at T=30K and 200K is plotted for 
different dopings. Naively, a decrease of spin-lattice relaxation is expected with reduced spin fluctuation as doping increases. However, the $1/T_1$ is enhanced by 
almost a factor of three at x=1.0 compared to lower dopings, which is rather a surprising result. Our result can not be understood if the magnetic
fluctuations are mainly driven by the interband scattering since the electron pockets disappear at x=1.0 \cite{Ding2}. In contrast, a pseudogap behavior and a modified 
Korringa relation have been reported in overdoped electron-type compounds \cite{Ning2, Ning3}. It seems that the spin fluctuations are stronger in our hole-doped 
compounds.

 The enhancement of the SLRR and the non-Korringa temperature dependence in KFe$_2$As$_2$ suggest that a new type of spin fluctuations develops at x=1.0. Since no long 
range order is observed in our crystals, such behaviors strongly suggest that the x=1.0 system is a paramagnetic phase approaching toward a second magnetic ordering, with 
the magnetic quantum critical point at a higher hole doping. For comparison, first-principles calculations proposed a new antiferromagnetic quantum critical point but 
with a lower critical doping $x_c\approx 0.8$, \cite{Fang_EPL_84}. It is also worthwhile to mention that the FeAs single crystals, although with a different lattice 
structure, has a helimagnetic order \cite{FeAs}. More work is necessary to examine the magnetic properties of KFe$_2$As$_2$ and also possibly more hole-doped compounds.

Next we discuss the superconducting properties of overdoped samples. We first analyze the SLRR of x=0.7 sample with $\mu _0H=12T//c$. Under this field, the Tc is 
suppressed to 16K and a few features are seen in the SLRR.  (i) The $1/T_1$ drops below Tc, and the Hebel-Slichter coherence peak is not seen in our crystal. Although 
$1/T_1$ is slightly enhanced at $T_C$, we have verified that the enhancement is due to a vortex dynamics effect
because the enhancement strongly depends on the field strength and orientation. (ii) The $1/T_1$ data has a prominent step-like feature at T$\approx$Tc/2 (8.5K). This 
feature clearly supports  a two-gap superconductivity. If we consider a two-gap system, the SLRR of a superconductor below T$_C$
can be written as,
\newline \newline $1/T_1\propto \displaystyle{\sum _{i,j=1,2}}  \int^\infty_0 n_{i}(E)n_j(E)f(E)(1-f(E))\, dE $
\newline \newline where $n_i$ is the superfluid density on two Fermi surfaces with different gaps, and $f(E)$ is the Fermi distribution function.
As observed in APRES\cite{Ding2}, two hole pockets at $\Gamma$ points have different gap values in optimally doped samples.  In the overdoped region,
the hole pocket with a  relative smaller superconducting gap can be quite large and contributes more carrier density.
 In this case, the contribution from the hole pocket with the smaller gap to $1/T_1$ in overdoped region is  enhanced and results in a step-like increase of the SLRR,
 compared to that in underdoped region where a less prominent step feature has been reported in LaFeAsO$_{0.92}$F$_{0.08}$ \cite{zheng1} and 
Ba$_{0.7}$K$_{0.3}$Fe$_2$As$_2$ \cite{Zheng2}. The reported SLRR by NQR studies on the powdered x=1.0 sample does not have such a step feature \cite{Fukazawa1}. It could 
be the gap at the large hole pocket is too small to produce such a feature. So far, we did not perform NMR studies on KFe$_2$As$_2$ below T$_C$. It would be worthwhile to 
check if the step feature can be seen below Tc by NQR or NMR measurements on KFe$_2$As$_2$ single crystals.

\begin{figure}
\includegraphics[width=6cm, height=5cm]{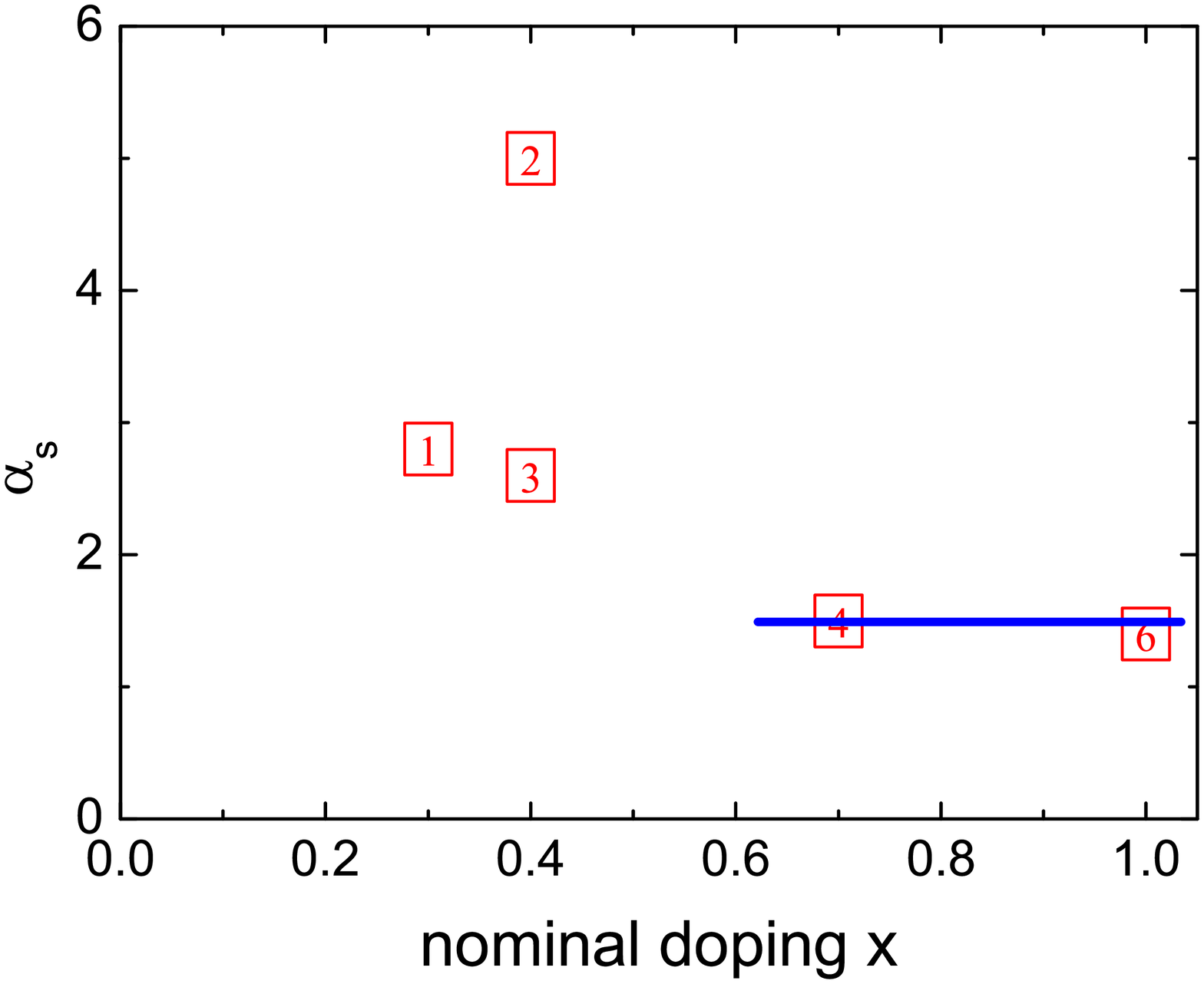}
\caption{\label{alpha}(Color Online) The power-law exponent $\alpha _s$ of $1/T_1\sim T ^{\alpha_s}$ of 1: x=0.3 (Ref. \cite{Zheng2}), 2: x=0.4 (Ref. \cite{Yashima1}), 3: 
x=0.4 (Ref. \cite{Fukazawa2}),  4: x=0.7 (current work), and 6: x=1.0 (NQR, Ref. \cite{Fukazawa1}) Ba$_{1-x}$K$_{x}$Fe$_2$As$_2$ samples.}
\end{figure}

Importantly, below Tc/2, the $1/T_1$ shows a power-law scaling with temperature, $1/T_1\approx T^{\alpha _s}$.  The scaling exponent $\alpha _s \approx 1.5$ holds down to 
our lowest temperature 1.5K. The reported KFe$_2$As$_2$ samples \cite{Fukazawa1} shows a similar power-law exponent $\alpha _S =1.4-1.5$ below Tc. Therefore, the 
power-law exponent $\alpha _S$ is quite universal over 30$\%$ range of doping concentration. In contrast, the power-law exponent is much larger and non-universal in the 
underdoped samples. In Fig.~\ref{alpha}, we plotted the  power law exponents in samples with different doping levels  reported by different groups. Below the optimal 
doping (x$\le$0.4), the power-law exponent ranges from 5 down to 2.8.

The small power-law exponent in the overdoped samples indicates strong low-energy excitations in the superconducting state, relative to those at low dopings. A strong low 
energy excitations has been observed in a La$_{0.87}$Ca$_{0.13}$FePO system \cite{Nakai}, although the cause is still unclear.  All our spin recovery is fitted nicely by 
a single T$_1$ component across the whole spectrum, which indicates that our SLRR data are intrinsic properties of a high quality sample. There are two possible origins 
of strong low-energy excitations in the overdoped samples. First, it is possible that the superconducting gap structure in the overdoped region is different from that in 
the underdoped region. In this case, the superconducting  pairing symmetry can be changed as indicated in both strong and weak coupling theories\cite{Singh1,seo2008}. 
ARPES and other experiments below Tc in the overdoped samples will be useful to address this issue in the future. Second,  even if the pairing symmetry is not changed as
doping increases, the effect of spin fluctuations on low energy excitations can be quite different in the overdoped region. Our normal state SLRR suggested a new type of 
spin fluctuations in the overdoped side. These spin fluctuations may produce q-dependent excitations and  lead to nodal-like behavior even in the s$\pm$ pairing state. In 
the overdoped region, since the size of the hole pocket is rather large, such a nodal-like behavior is more likely to take place. Under such circumstances, even in the 
s$\pm$ pairing state, the SLRR can be described by a dirty s-wave or a d-wave superconductor \cite{Bang_PRB, seo2009a}.

In summary, our study of the doping and the temperature dependence of the spin-lattice relaxation rate in heavily overdoped Ba$_{1-x}$K$_{x}$Fe$_2$As$_2$ suggest that a 
new type of spin fluctuations develops as the doping close to x=1.0. These results strongly suggest that system evolves towards a new magnetic quantum critical point, and 
rule out a simple Fermi-liquid description of the normal state. In the superconducting state, no coherence peak is seen, and a step-like behavior in the SLRR at Tc/2 
supports the idea of a two-gap superconductivity. The SLRR below T$_C$ is characterized by a universal power-law behavior $1/T_1\sim T^{1.5}$, which suggests strong 
low-energy excitations below T$_C$, in contrast to the behavior in the under- or optimally doped samples. Such low-energy excitation may be correlated with a different 
pairing symmetry or the new type of spin fluctuations in the overdoped side.

W.Y. and G.F.C. are supported by the National Basic Research Program of China. W.Y. acknowledges the discussions with S. E. Brown. J.P.H. acknowledges the support from 
the Institute of Physics, Chinese Academy of Sciences. Research at McMaster is supported by NSERC and CIFAR.


\end{document}